# Simulation toolkits at the molecular scale for trans-scale thermal signaling

Ikuo Kurisaki[1*] and Madoka Suzuki[2*]

[1] Waseda Research Institute for Science and Engineering, Waseda University, Bldg. No.55, S Tower, 4th Floor, 3-4-1 Okubo Shinjuku-ku, Tokyo 169-8555, Japan

[2] Institute for Protein Research, Osaka University, 3-2 Yamadaoka, Suita, Osaka 565-0871, Japan

* Corresponding authors.

*E-mail addresses:* ikuo.kurisaki@aoni.waseda.jp (I. Kurisaki),

suzu_mado@protein.osaka-u.ac.jp (M. Suzuki)

# Abstract

Thermogenesis is a physiological activity of releasing heat that originates from intracellular biochemical reactions. Recent experimental studies discovered that externally applied heat changes intracellular signaling locally, resulting in global changes in cell morphology and signaling. Therefore, we hypothesize an inevitable contribution of thermogenesis in modulating biological system functions throughout the spatial scales from molecules to individual organisms. One key issue examining the hypothesis, namely, the "trans-scale thermal signaling," resides at the molecular scale on the amount of heat released via individual reactions and by which mechanism the heat is employed for cellular function operations. This review introduces atomistic simulation tool kits for studying the mechanisms of thermal signaling processes at the molecular scale that even state-of-the-art experimental methodologies of today are hardly accessible. We consider biological processes and biomolecules as potential heat sources in cells, such as ATP/GTP hydrolysis and multiple biopolymer complex formation and disassembly. Microscopic heat release could be related to mesoscopic processes via thermal conductivity and thermal conductance. Additionally, theoretical simulations to estimate these thermal properties in biological membranes and proteins are introduced. Finally, we envisage the future direction of this research field.

# Contents

# 1. Introduction

Thermogenesis is a physiological activity of internally releasing heat. It is widely observed among living organisms; not only in mammals and birds but also in insects and plants. Physiological roles of thermogenesis include maintaining the temperature of the body higher than that of a cold environment or attracting insects for pollination. Thermogenesis originates from biochemical reactions in cells [1,2]. Shivering thermogenesis is a thermogenic process by myosin ATPase in skeletal muscles in animals. The other mechanism that does not accompany muscle contraction is called non-shivering thermogenesis, such as those by the uncoupling protein 1 located in the mitochondrial inner membrane, or by SERCA, which is a P-type ATPase at the membrane of the sarcoplasmic reticulum that is a $Ca^{2+}$ store in muscle cells. These biochemical reactions often develop to enthalpically stabilize the system, and usually, the reactions take place in specific sites or organelles in a cell such as the mitochondria or sarcoplasmic reticulum. A variety of cellular responses to a steep temperature gradient and heat pulses that we and other authors have discovered [3] suggest that those cellular hot spots could inevitably contribute to intracellular signaling locally, resulting in global changes in cell morphology and signalings such as the ultrafast neurite growth [4], muscle contractions [5,6], and the uncontrolled thermogenesis in malignant hyperthermia [7]. Therefore, we hypothesized the contribution of thermogenesis throughout the spatial scales in the hierarchy of a biological system from molecules, cells, tissues, and individual organisms as the "trans-scale thermal signaling."

One key issue in the trans-scale thermal signaling hypothesis resides at the molecular scale to understand the amount of heat released via biochemical reactions and the mechanism by which the heat is employed for cellular function operations. This kind of question is hardly accessible even by state-of-the-art experimental methodologies of today. Under such circumstances, theoretical simulation can be a feasible technical counterpart to examine atomistic characterization of the phenomena and to provide seamless and comprehensive descriptions over the scales (**Figure 1**).

This review introduces simulation tool kits for studying the mechanisms of thermal signaling processes at the molecular scale, including conventional, and more recently developed ones. We discuss future developments of these approaches and expected collaboration with experimental observations recalling the applicability of theoretical approaches to realistic systems.

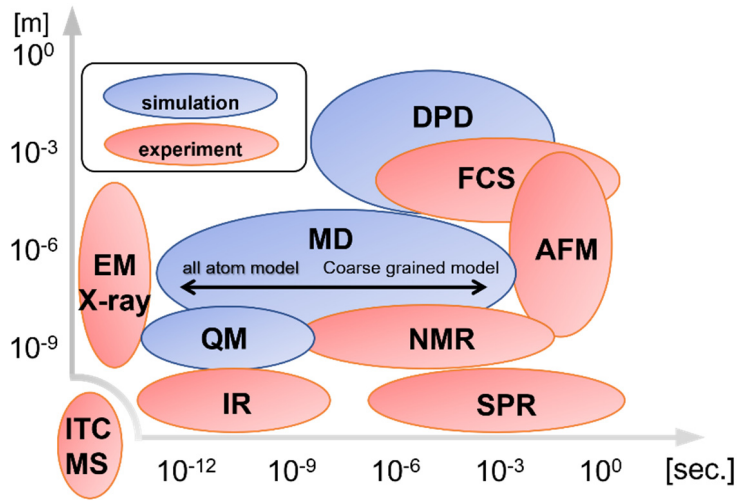

**Fig. 1.** Temporal and spatial scales are accessible by either computer simulation (blue) or experimental (red) methods. MS: mass spectroscopy, ITC: isothermal titration calorimetry, IR: infrared spectroscopy, SPR: surface plasmon resonance, EM: electron microscopy, X-ray: X-ray crystallography, NMR: nuclear magnetic resonance, AFM: atomic force microscopy, FCS: fluorescence correlation spectroscopy, QM: quantum mechanics, MD: molecular dynamics, DPD: dispersion particle dynamics.

## 2. Heat release via a biochemical reaction

The physicochemical entity of life is a network of cascades in biochemical reactions in cells. The monumental number of biochemical reactions concertedly progress to sustain the living conditions of the cell [8,9]. Among several biochemical reactions, ATP, and GTP hydrolysises have been extensively discussed as essential chemical reactions. The ATPase [10-13] and GTPase [14-17] regulate critical biological processes such as cell motility and signal transduction by converting the chemical energy of ATP and GTP, respectively, into mechanical work via the reactions. This section discusses the applications of quantum mechanics simulations to analyze reaction mechanisms of ATP/GTP hydrolysis and classical molecular dynamics simulations to examine the time course of post-reaction processes.

**2.1 *Ab initio* approaches for biochemical reactions with quantum mechanics simulation**

From an atomistic point of view, a chemical reaction is characterized by rearrangements, formations, and breaks of chemical bonds via changes in electron distributions around atomic nuclei. Therefore, the physicochemical mechanism can in principle be examined in the framework of quantum mechanics (QM) theory, which explicitly treats the distributions of electrons in the system considered. The governing equation of QM is the Schrödinger equation. We can only obtain the solutions using numerical approximation techniques, such as molecular orbital (MO) theory and density functional theory, for atomistic systems with multiple electrons. However, such numerical approaches include a time-consuming process and diagonalization of a large sparse matrix. Additionally, computational cost dramatically increases in the proportion of (at least) third power of the number of electrons in the system, even if we use a relatively simple method, such as the Hartree-Fock equation [18], which is the most elementary *ab initio* calculation method based on MO theory.

Thus, QM/molecular mechanics (MM) simulations are usually employed when we study a huge system with >1000 atoms such as a biomolecule in an aqueous solution. Only atoms essential for chemical reactions are considered for QM calculations in the QM/MM scheme while the remaining is treated using MM calculations with empirical force field parameters. This hybrid approach significantly reduces computational costs to the level where realistic computational resources can examine mechanisms of chemical reactions in the huge system. The QM/MM simulation scheme was first proposed by Warshel and Levitt in 1976 [19]. It has been extensively employed to examine the physicochemical mechanisms of biochemical reactions at the microscopic scale. The results allow (1) to examine the conformation of intermediate states of the reaction, (2) to analyze physicochemical changes such as charge redistributions, formations breaks of atomic bonds, and (de)protonations, (3) to identify a possible reaction pathway within a realistic time frame, and (4) to estimate atomic heat released through the reaction in terms of enthalpy because QM/MM simulations provide an energy landscape of the biochemical reaction and a full set of atomic coordinates of the reaction system.

A monumental number of QM/MM studies were conducted on biochemical reactions [20-24]. Among those, the ATP hydrolysis mechanism by a cytoskeletal protein, actin [25], and a molecular motor, myosin II [26,27], appears as one of the case studies within the scope of the current review. The enthalpy change through the reaction is approximately 12 kcal/mol in actin according to QM/MM studies. This value is similar to the enthalpy that is experimentally derived, which is 10 kcal/mol. Two possible ATP hydrolysis mechanisms have been examined for the study on myosin II, i.e., a nucleophilic attack of one water molecule and a proton transfer involving two water molecules. The study concluded that the latter mechanism can progress due to its lower activation barrier of 10 kcal/mol compared to that of the former of 20 kcal/mol within the biological time frame. Again, the simulated value is comparable to the heat release obtained from experimental studies [27], thereby supporting the reliability of QM/MM simulations.

At present, QM/MM simulation scheme can elucidate plausible ATP hydrolysis mechanisms inside the cell at the atomic level and the total amount of heat released from the reaction. Our research interest may take a step forward to examine the possibility that the heat released from the hydrolysis reaction drives a biomacromolecule to express its function. Hence, atomistic dynamics need to be investigated in the biochemical reaction systems. However, QM simulation requires relatively large computational costs even if we employ the hybrid, QM/MM scheme. Therefore, simulating temporal changes of chemical reaction in the frame of the QM/MM method is infeasible by numerically solving the Newtonian equation of motion with QM/MM-derived energies. An alternative approach is required to obtain a dynamic view of the biochemical reaction processes at the atomic level.

## 2.2 Effective treatment of chemical conversion in the framework of classical molecular dynamics simulation

Atomic force microscopy and time-resolved X-ray crystallography allow direct molecular motion monitoring at a near-nanometer scale [28-30]. However, tracking transient processes remained challenging, which progress with the order of *pico* or *femto* second even for state-of-the-art experimental techniques, with a spatial resolution of sub-nanometer. Time-resolved ultraviolet resonance Raman spectroscopy is one of the powerful experimental tools to study the vibrational energy flow, i.e., thermal diffusion, in proteins. [31] Meanwhile, it does not give the information in individual molecules but only the ensemble average obtained from protein solutions, nor quantitative information for local energy change in proteins so far. Atomistic molecular dynamics simulation is their technical counterpart because of its finer temporal resolution combined with atomic resolution. Meanwhile, it should be noted that chemical reactions remain '*rare events*' for conventional atomistic simulations. The reaction times of ATPase and GTPase are usually 1000-fold or longer than the time scale of simulation that is accessible with currently available computational resources (**Fig. 2**).

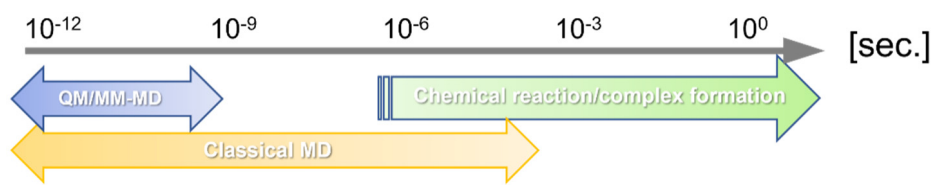

**Fig. 2**. Temporal scales accessible by quantum mechanics/molecular mechanics-molecular dynamics (QM/MM-MD) and classical molecular dynamics (MD) methods, as shown in blue, and orange, respectively. The temporal scales associated with chemical reactions and complex formation are colored green.

Under such a circumstance, we developed a simulation approach to effectively examine the consequence of chemical reactions, i.e., the chemical conversion of reactant molecules into product ones, in the framework of classical molecular dynamics [32]. The main idea of this approach is a switchover of force field parameters between the reactant and product systems [33,34], so we call it the switching force field molecular dynamics (SF2MD). SF2MD was applied to test John Ross' conjecture which remarks microscopic mechanism of conversion between chemical energy and mechanical work via ATP/GTP hydrolysis in enzymes (**Fig. 3A**) [35]. We found that the physicochemical process discussed in Ross' conjecture appears in the atomistic simulation of the GTP hydrolysis reaction of the Ras-GTP-GAP system with SF2MD (**Fig. 3B**) [32]. Repulsive Coulomb interaction energy acting between GDP and $P_i$, which is the main component of non-bonded energy, was converted into kinetic energy of these reaction products (**Fig. 3C**). However, the directionality of velocity was rapidly lost through thermalization within 1 ps following $GDPP_i$ generation. Therefore, the energy conversion cannot be a source of mechanical work to promote chemical processes such as Ras-GAP dissociation and conformational change of Ras. These observations challenge the remark in Ross' conjecture.

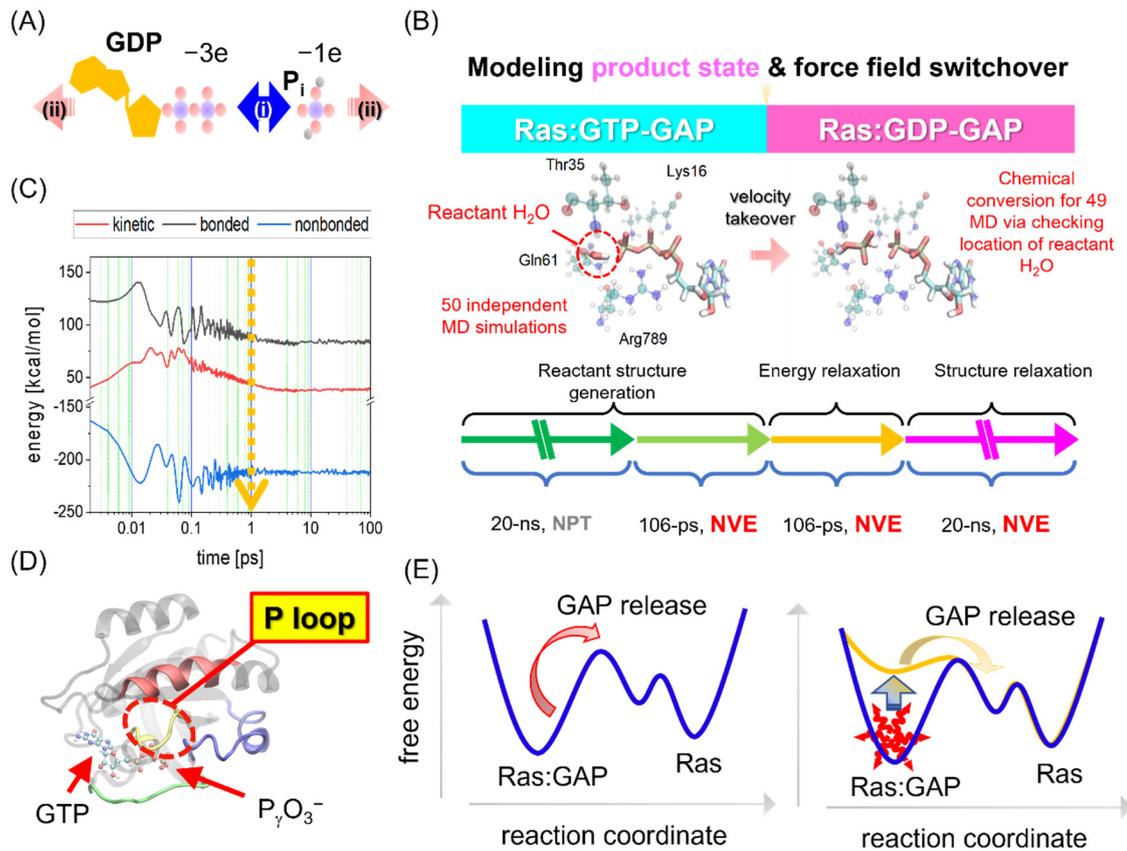

**Fig. 3**. Outline of atomistic simulation study with the SF2MD approach. (A) Schematic illustration of Ross' conjecture; the mechanism of generating mechanical work via ATP/GTP hydrolysis reaction, as proposed by John Ross. (i) Repulsive Coulomb interaction acting between reactants (GDP and inorganic phosphate, $P_i$, in this panel) is converted into (ii) *directional* kinetic energy and does the mechanical work on the system upon the hydrolysis. (B) Simulation protocol of SF2MD for Ras:GTP-GAP complex system. NPT denotes physical ensemble under constant particle number, pressure, and temperature condition, while NVE denotes physical ensemble under constant particle number, volume, and temperature. (C) Energy relaxation process after GDP-inorganic phosphate generation in SF2MD for Ras:GTP-GAP complex system. Red, black, and blue lines denote kinetic, bonded, and non-bonded energies, respectively. The orange arrow indicates the time point of 1 picosecond. (D) Schematic illustration of Ras protein. Ras protein and GTP are shown with ribbon and ball and stick, respectively. $P_gO_3^-$ denotes an inorganic phosphate at the g position in GTP. Helix-C, P-loop, Switch-I, and Switch-II (functional regions in Ras protein) are colored red, yellow, green, and blue, respectively. (E). Conventional (left) and proposed mechanism (right) of GTP hydrolysis accelerating Ras:GAP dissociation to step forward to the next biological condition.

Furthermore, analyses of SF2MD-derived atomic trajectories revealed an unprecedented change in Ras protein. The phosphate-binding loop (P-loop) in Ras protein (**Fig. 3D**) gained approximately 5 kcal/mol of potential energy (almost half of the free energy derived from ATP/GTP hydrolysis). This result indicates that the mechanical work generated via the GTP hydrolysis is substantially stored in the P-loop as a form of local interaction energy. Ross' conjecture highlighted their transient collision of GDP and $P_i$ to surrounding protein atoms as a source of mechanical force generation. Conversely, SF2MD-based atomistic simulation demonstrated that such mechanical processes are not important for generating mechanical work. Rather, the hydrolysis of triphosphate nucleotide does mechanical work by producing emergent steric interaction at one domain of the enzyme and relaxation at the other site; namely, a shift of biomolecular system to a non-equilibrium state and reshaping of the potential energy landscape (**Fig. 3E**). There remains an open question on how enzymatic ATP/GTP hydrolysis generates mechanical works.

Our SF2MD simulations, as discussed in **Figure 3C**, can track atomic velocities, thereby being feasible to examine the atomistic process of the heat release via a chemical reaction although the heat released through the chemical change from GTP to GDP and $P_i$ was not the main focus of the above-mentioned study. The technical advantage of SF2MD is that it allows for quantitative analysis of the apparent heat release mechanism, which is based on the conversion between potential and kinetic energies via the change in chemical species. This approach is in contrast with conventional approaches, with which the heat release is considered as randomly assigned atomic velocities [36]. The SF2MD is expected to be an important bottom-up approach for comprehensively understanding the molecular thermogenesis in living cells at the atomic level.

## 3. Heat release via multimeric biopolymer complex formation

The function of biological nanomachines consisting of multiple biopolymers, such as protein, DNA, and RNA, is regulated by multimeric biopolymer complex formation and their disassembly [37-39], where the rearrangement of non-bonded, inter-, and intra-molecular interactions play essential roles. These nanomachines regulate critical biological processes, such as cell signaling [40-42], DNA repair [43,44], protein synthesis [45,46], and RNA editing [47-49]. Ordered formation and disassembly processes of multimeric biopolymer complex could be alternative sources of heat release in the cell, recalling that a number of those processes frequently occurred in living cells. This section remarks on theoretical approaches to examine such complicated processes involved with multiple biopolymers.

**3.1 Experimental and computational prediction of the formation order of multimeric protein complex**

From a physicochemical point of view, the formation and disassembly processes of multimeric biopolymer complex can be characterized by (1) a set of intermediate states defined with subcomplexes that are formed during the processes and by (2) physicochemical mechanisms of transitions between the intermediate states. Mass spectrometry (MS) analysis has become a powerful tool to systematically identify a set of subcomplexes emerging during multimeric biopolymer complex formation or disassembly [50-60]. Biopolymers are excessively charged and then detected as signals of a set of subcomplexes in MS (**Fig. 4**). Molecules that bind each other more weakly tend to be promptly dissociated from the multimeric biopolymer complex. We can investigate which subcomplex population appears in the earlier phase of the disassembly process by systematically changing the total amount of excessive charges added to biopolymer samples. The information of the sets of subcomplexes is then employed to deduce the dissociation order of subcomplex in the multimeric biopolymer complex. The order of disassembly is often explained by the relative strength of interactions that act between the biopolymers. In other words, we could deduce a possible set of subcomplexes that appear during formation and disassembly processes at the molecular level by analyzing the relative strength between interfaces of individual biopolymers with knowledge of multimeric complexes. Salt-bridge formation between subunits is one example of descriptors for the relative strength of the inter-molecule interaction [54].

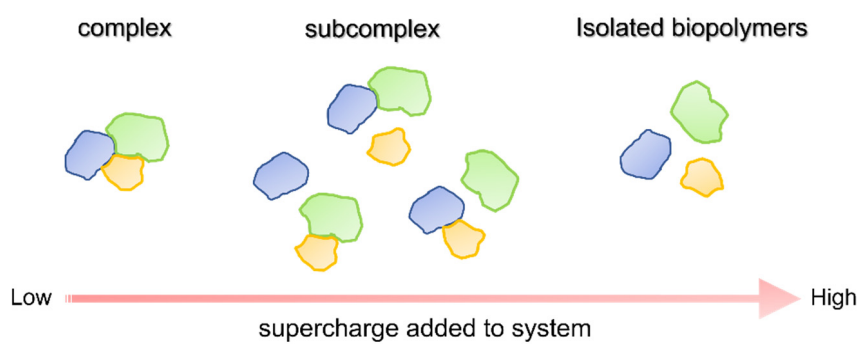

**Fig. 4.** Biopolymer complex leads to subcomplexes by supercharging samples.

This kind of descriptor can be obtained by analyzing the atomistic coordinates of multimeric complexes. Meanwhile, a disassembly of such complexes could take a reverse route for the multimer complex formation under specific conditions [51]. This assumption of multimeric complex formation-disassembly reversal processes and the relationship between the order of disassembly and the strength of inter-subunit interactions will encourage to deduce the order of disassembly from the atomistic structures of multimeric biopolymer complexes without performing MS experiments.

The prediction of the multimeric protein complex pathway has been extensively studied during the last two decades in the field of structural bioinformatics [61,62]. Path-LZerD by Kihara et al. now becomes an outperforming method among the techniques developed for this purpose [63,64]. Earlier studies on subcomplex formation order prediction employed buried surface area [51,55] as a rough estimation of the binding free energy, while such a quantity cannot be calculated without *a priori* knowledge of overall complex structures consisting of multimeric biopolymers. Path-LZerD is designed to construct multimeric protein complexes from a set of monomeric proteins without *a priori* knowledge of overall complex structures. This technical advantage is realized by the use of knowledge-based potentials to evaluate binding preferences between subunits. Thus, Path-LZerD is a more universal approach than previous methods.

At present, the identification of subcomplexes involved in multimer complex formation and the disassembly processes as its reverse process have become technically feasible. However, the complementary usage of these two proteomic approaches, MS, and structural bioinformatics, such as Path-LZerD, remains unsatisfactory to elucidate the formation mechanisms of multimeric protein complexes under realistic cellular environments. Complex formation of biopolymers and their disassembly progress in a certain period and specific biological conditions. MS requires supercharged biopolymers even if the observed samples are prepared under pseudo-physiological conditions as in the native MS, which is a more recently developed experimental technique. Atomistically considering the effects of solvation and of interaction with co-solvent, such as ions, and metabolites, namely, molecular crowders, of which effects are fairly important to understand the formation mechanisms of biopolymer complexes and their functions, remain challenging for bioinformatics approaches [30,65,66]. The interaction between biopolymers and the molecular crowding milieus is nonspecific and relatively weak; hence, the atomic coordinates of the interaction are usually missing in experimentally-determined biomolecule structures that structural bioinformatics employs.

**3.2 Hybrid Monte Carlo/molecular dynamics simulations to predict disassembly order**

Under these circumstances, we recently proposed a hybrid Monte Carlo/molecular dynamics (hMC/MD) approach [67] and applied it to simulate the disassembly process of serum amyloid P component (SAP) homo pentamer, which has been employed as a benchmark of MS experiments [53,68]. Dissociation would happen within the timescale of sub-milli seconds (**Fig. 2**), thus they are rarely observed through atomistic molecular dynamics simulation with realistic computational resources.

Such a "slow" dissociation process is accelerated by superposing external force on a pair of subunits of SAP pentamer in hMC/MD. The atomic structure of a multimeric protein complex is randomly selected at each simulation cycle to run the simulation scheme with minimum *a priori* knowledge for disassembly, while "accept" or "reject" of the partly dissociating structure generated is judged using Monte Carlo simulation. This probabilistic step is employed to exclude highly anomalous and unstable atomic structures that often appear during such simulations which accelerate monomer-pair dissociation. The complete hMC/MD simulation scheme is illustrated in **Figure 5A**. The results obtained by hMC/MD were consistent with the observations derived from the MS and structural bioinformatics approaches reported earlier by Hall [54] concerning the emergence order of subcomplexes in the pentamer disassembly processes (**Fig. 5B**). Additionally, the hMC/MD simulations predict that the ring shape of SAP pentamer undergoes an opening process before the dissociation of subunits.

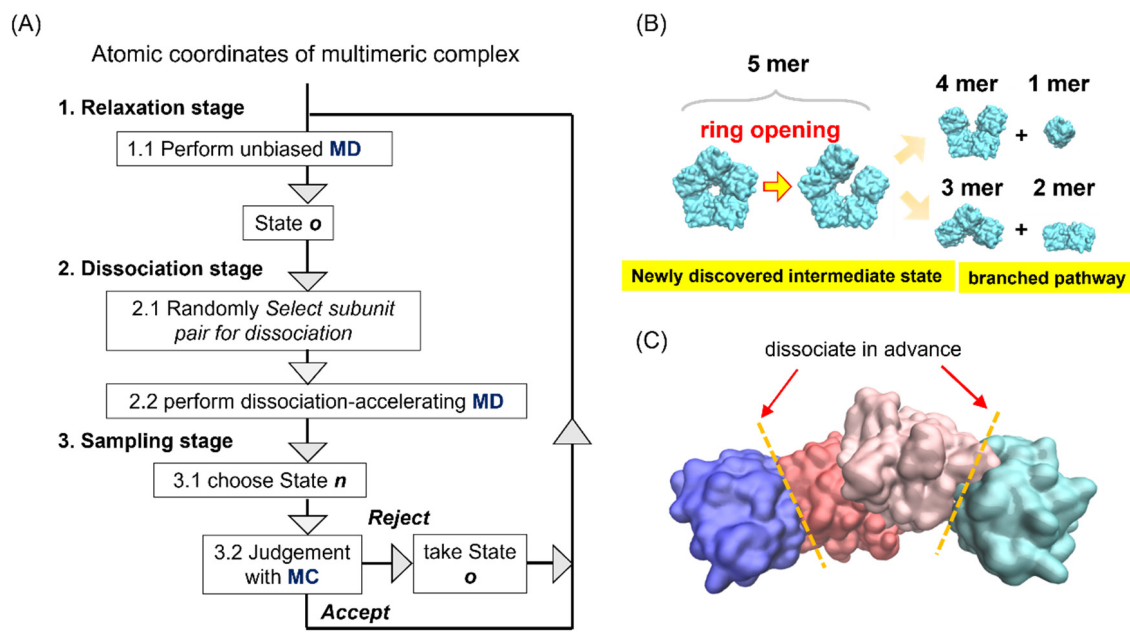

**Fig. 5**. (A) Schematic chart of hybrid MC/MD simulation for disassembly of the multimeric protein complex. The simulation cycle is repeated by several hundred times to completely disassemble a multimeric complex into subunits. (B) Theoretical predictions of disassembly order of multimeric proteins, serum amyloid P component (SAP) home pentamer. There are two different dissociation pathways: a pair of tetramer and monomer or trimer and dimer. (C) Tryptophan synthase hetero tetramer. α and β subunits are illustrated with blue and red surfaces, respectively. Orange dotted lines illustrate the interaction interfaces where the pair of monomers is dissociated in advance.

Analysis of the free energy for this new process demonstrates a possibility that one can observe the ring-opening of SAP pentamer within experimental time scales; hence, the process should be experimentally examined by methods such as fluorescence spectroscopy or atomic force microscopy. Additionally, the multimer formation process could be similarly investigated if the assembly was approximately regarded as a reverse process of disassembly [51]. Thus, the formation and disassembly process of SAP pentamer could be suitable to examine atomistic mechanisms of multimeric biomolecular complex formation and to bridge the gap between simulations and experiments.

The hMC/MD simulation scheme would have the potential to be a general methodology to examine the disassembly processes of multimeric protein complexes. Therefore, we further extended the scheme to heteromeric protein complexes. We considered a selection bias of dissociating monomer pairs based on a quantitative metric of inter-protein interactions [69]. Prediction of the disassembly process of heteromeric multimers is more challenging than that of homomeric multimers. This is because the diversity of the protein-protein binding interface critically influences the dissociation order among monomer pairs. However, our simulations show a >85% success rate to predict the disassembly order of heteromeric multimers, such as tryptophan synthase tetramer (**Fig. 5C**) and acetyl-CoA carboxylase carboxyltransferase tetramer. The success rate is much higher than 33% which is calculated by randomly selecting a dissociated pair (a detailed discussion is given in the related work [69]).

These achievements indicate that the hMC/MD approach is possible to provide a complete set of atomic coordinates for multimeric protein complex systems at any of the intermediate states in disassembly processes. This information is primary knowledge of the physicochemical analyses of reaction kinetics. We can calculate the free energy profiles of disassembly processes using the hMC/MD-derived atomistic trajectories for additional umbrella sampling MD simulations, for example [67]. It allows estimating the amount of heat released via the dissociation of each subunit pair as a difference of the free energy.

Weighted ensemble (WE) simulations [70-72] and parallel cascade (PaCS) simulations [73-75] can also be applied to estimate kinetic rate constants of the subprocesses in multimeric biomacromolecule disassembly. Simulating a complete route of disassembly of multimeric complexes remains challenging although these simulations have successfully estimated kinetic rate constants with high accuracy that are comparable with those obtained from biochemical measurements. WE and PaCS simulations repeatedly generate a substantial number of short MD simulations, each length of which is up to 100 ps so far, to accelerate rare events such as protein complex dissociation. However, system changes that happen in such short simulations are promptly relaxed and cannot be sustained under thermal fluctuation if a chemical process has several change directions, as in the case of multimer disassembly (see discussion in Supporting Information of the study in which we first proposed hMC/MD approach [67]). Meanwhile, our hMC/MD approach itself simply provides atomistic coordinates of disassembly processes of multimeric biopolymer complex. Therefore, additional simulations are necessary to examine the physicochemical property of the disassembly process such as the free energy profiles of each dimer dissociation. We could reliably estimate the thermodynamic and kinetic properties of each dimer dissociation process consisting of a disassembly process using hMC/MD-derived knowledge of disassembly order for PaCS or WE simulations.

Our hMC/MD simulation, together with WE, and PaCS, can be the third methodology in the structural proteomics research field due to the complementary roles for the two pre-existing methodologies, MS, and structural bioinformatics approaches (**Table 1**). We hypothesize that the combined uses of the hMC/MD approach and these established simulation techniques will give insights into the time scale of heat release processes, as well as the total amount of heat released via multimeric protein complex (dis)assembly.

**Table 1**. Accessible information with methods to examine multimeric biopolymer complex formation and disassembly.

| Parameters of Interest | Experiments | Theoretical Computation | | |
|---|---|---|---|---|
| | | Structural bioinformatics | Atomistic simulation | |
| | Mass spectroscopy | | Hybrid Monte Carlo/molecular dynamics | Weighted ensemble, parallel cascade sampling |
| Spatial resolution | molecule | atom | atom | atom |
| Temporal resolution | N.A. | N.A. | femtosecond or greater, dependent on available computational resource | femtosecond or greater, dependent on available computational resource |
| Atomic coordinate | N.A. | subcomplex appearing in pathway, in ignorance of the coordinates of dissociated monomers | any molecules | any molecules |
| Physicochemical property | relative population of subcomplexes | anything calculated from atomic coordinates, approximate | anything calculated from atomic trajectories, using additional structure sampling | anything calculated from atomic trajectories, exact in the computational framework |
| Multimer size | any size | any size | any size | dimer, at most so far |

## 4. Estimation of thermal properties of biopolymers

Temperature gradients inside a cell cause heat transfer between biological molecules locally and may generate convective heat transfer in the cytoplasm globally [76]. These physical processes can be supposed as the microscopic and mesoscopic entities of thermal signaling in the cell. The thermal conductivity of intracellular "materials" and thermal conductance between the materials with different thermal conductivity values are essential physical quantities to characterize the mechanism of such thermal signaling because the spatial and temporal dynamics of these processes are described by the heat diffusion equation and hydrodynamic equation.

Interestingly, the values of thermal conductivity of proteins have been found around one-third of that of water. Intracellular space is crowded with a variety of biomolecules [77]. Meanwhile, the spatial distribution of such biomolecules is inhomogeneous, and it changes over time. Reliably and comprehensively characterizing the thermal properties inside the cell by experimental means remains challenging for such complicated molecular systems [78-80]. Theoretical simulations are thus again expected as technical counterparts. This section discusses simulation techniques that have been developed to examine the thermal properties of biomolecules and their applications in biomacromolecules such as lipid membranes and proteins.

**4.1 Biological membranes**

Simulation techniques for lipid membranes have been developed in the field of material science [81-83]. A heat flow can be generated by giving high and low heat baths in a given molecular system. Several simulation studies have already reported the atomistic origin of thermal diffusion processes and the functional roles of lipid membranes. For example, Nakano et al. employed non-equilibrium simulation approaches to estimate thermal conductivity and thermal conductance between dipalmitoylphosphatidylcholine (DPPC) lipid membranes and water [82]. They demonstrated an anisotropy of thermal conductivities in lipid membranes by dissecting the mechanical contributions of the head group of lipids, tail group of lipids, and surrounding water. Youssefian et al. focused on the thermal property of DPPC lipid membrane at various temperatures and revealed the lowest thermal conductivity around its fluid-gel phase transition temperature [83]. The phase transition of biological lipid membrane has been suggested to influence the functional expression of biomolecules, such as ion channels [84], thus such studies possibly open a new avenue to elucidate signal transduction in the cell. Their subsequent study examined differences in thermal properties among archaeols which are common components of the lipid membrane of archaea. Their results indicate the effects of asymmetry in the tails of lipids for thermal rectifiers, suggesting the biological significance of archaeal membranes to withstand extreme conditions [85].

The accumulated knowledge of biological membranes tells us the wide diversity of their compositions [86,87]. The compositions are dependent on cell and tissue types, and they often change gradually via aging processes [88,89]. Scientific merit might not be worth the experimental efforts to comprehensively examine the thermal properties of biological membranes for all possible biological conditions due to such complexities. Theoretical prediction should have a chance to draw attention to the field of biological membranes.

**4.2   Proteins**

Lipid membranes usually take a solid-like state in cells, thereby being accessible by conventional experimental and computational methods [85,90]. The experimental methods for examining the thermal properties of solid-like materials are designed to use amorphous solids or water-insoluble films of biomacromolecules [90] so that the same methods are not directly applicable under hydrated conditions where proteins often express their biochemical functions.

Leitner et al. have reported pioneering simulation studies on the theoretical estimation of the thermal conductance of a protein, where they employed the formulation of vibrational energy transfer [91,92]. Thermal conductance can be calculated under the theoretical framework once vibrational mode densities of biomolecules are given from either experimental or computational simulation methods. They systematically examined the dependence of thermal conductance on system temperature and found an increase in thermal conductance along the temperature increment. Their approach was applied to examine the effect of saccharides (glucose, galactose, sucrose, and trehalose) as molecular crowders on the thermal conductance of proteins. Their study revealed that trehalose molecules surrounding proteins significantly reduce thermal conductance between protein and water, hence the trehalose molecule functions as a thermal insulator for proteins [93].

Yamato et al. developed a simulation method to estimate the thermal conductivity of proteins in the framework of linear response theory [94]. The estimated values are compared to experimental values and earlier simulation methods. Notably, their method needs only atomistic simulation trajectories as input to calculate the thermal conductivity. Thermal conductivity can be estimated without additional *a priori* knowledge of physical quantity, such as heat capacity, whose value is technically difficult to determine for both experimental and computational methods. Moreover, their simulation scheme requires a substantial amount of computational resources and data storage.

Another pioneering simulation work was performed by Lervik et al. to calculate the thermal conductivity and conductance of proteins [95]. They executed a non-equilibrium MD simulation, where protein, and solvent molecules are coupled with high and low thermostats, respectively, to generate heat flow from hydrated protein to the surrounding solvent. Then, the data were fit by the theoretical solution of the heat diffusion equation to estimate the thermal conductivity and conductance of the protein (**Fig. 6**). Their simulation scheme is a well-designed combination of atomistic simulation and physical theory. The calculated results agreed well with known experimental data. Additionally, the theoretical scheme is relatively easy to implement by making an in-house script that fits an MD simulation-derived temperature relaxation profile on the numerical solution of the heat diffusion equation. Meanwhile, Lervik's simulation scheme may be improved for extended applications to various proteins with larger sizes such as biological nanomachines. From the viewpoint of computational cost, the scheme is inefficient to examine non-spherical proteins. They employed the water sphere to model a hydrated protein due to consistency with their usage of the spherical form of the heat diffusion equation. Proteins with anisotropic shape need a much greater number of water molecules than spherical proteins if we model the hydrated condition of such protein with a water sphere.

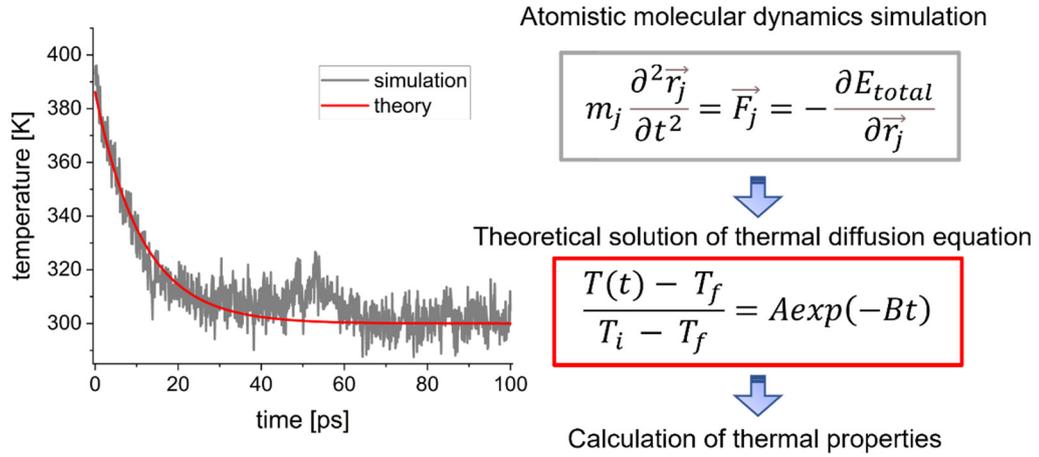

**Fig. 6**. Lervik's scheme to estimate thermal properties (thermal conductivity and conductance) by combining atomistic simulation and a physical model of thermal diffusion. $E_{total}$ is total energy of a biopolymer system. $r_j$ vector denotes an atomic coordinate of the $j^{th}$ atom. $T$ is assigned as temperature, where subscripts $i$, and $f$ mean initial and final, respectively.

Therefore, we addressed improving their simulation scheme for extended applications [96]. First, we performed similar non-equilibrium simulations under periodic boundary conditions for model proteins, myoglobin (Mb), and green fluorescent protein (GFP) which have been discussed in Lervik's study. The calculated values of thermal conductivity were comparable to those given by Lervik's study when we used a sufficiently large system size for the simulation. Lervik's simulation scheme was suggested to work with MD simulation trajectories obtained under periodic boundary conditions, which can be used to significantly reduce computational costs for proteins with arbitrary shapes. Meanwhile, thermal conductance appears to deviate from the values given in Lervik's study. This discrepancy is possibly due to the delicate nature of parameter fitting. The estimation of parameters for the numerical solution of the heat diffusion equation are sensitive to simulated temperature trajectories in the scheme that Lervik employed. However, once the sensitivity issue is cleared, their scheme is much more advantageous concerning relatively low computational cost and simple interface program implementation.

Next, we extended Lervik's simulation scheme by considering proteins with cylindrical shapes, as proteins and their complexes often demonstrate non-spherical shapes (**Fig 7**). Then, we solved the heat diffusion equation in the form of a cylinder with two different boundary conditions. The simulation led to systematically greater thermal conductivity and thermal conductance due to the theoretical formulation (detailed discussion is given in the original study [96]). This result does not simply indicate the inferiority of cylindrical shape models. We employed the united atom model with relatively simplified atomistic force field parameters (GROMOS96 43a2 force field [97]) to obtain MD trajectories of proteins under nonequilibrium conditions, as in the case of the original study by Lervik. Atomistic simulations with sophisticated force field parameters of today may reveal underestimated values of thermal properties of proteins with a spherical shape model and show the superiority of the cylindrical shape model.

This observation remains to be tested and left as a future challenge.

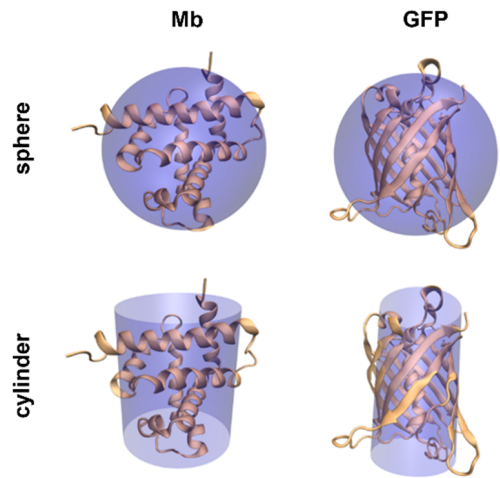

**Fig. 7**. Molecular structures of proteins (Mb: myoglobin; GFP: green fluorescence protein), overlapped with transparent sphere, and cylinder.

In summary, we introduced three theoretical approaches to estimate the thermal conductivity and conductance of proteins with inhomogeneous shapes. These approaches are still in their infancy. Further technical developments are required to estimate the thermal properties of a variety of proteins.

**Summary and Outlook**

This review highlights the simulation tool kits that are applicable for studying the microscopic mechanisms of the trans-scale thermal signaling processes based on MD.

The first biological topic was the heat release in the cell that is often derived from biochemical reactions such as ATP/GTP hydrolysis. Quantum mechanics simulations are critical tools to clarify the reaction mechanisms of ATP/GTP hydrolysis. Meanwhile, the classical MD simulations that effectively consider chemical conversion, called SF2MD, examine the time course of post-reaction processes, such as heat release into the whole system and structural rearrangement of biopolymer. We could examine elementary processes of heat release and give physicochemical insights into thermal signaling from an atomistic point of view using these simulation techniques. Next, we discussed the formation, and disassembly of multimeric biopolymer complexes as another source of heat release in the cell. Such a complicated process can be examined by combining MS, BS, and atomistic simulations with hMC/MD approach. The validity of simulated trajectories is evaluated by considering MS and BS observations. We could estimate the amount of the released heat through the rearrangement of the multimeric biopolymer complex under physiological conditions by performing further atomistic simulations to calculate free energy profiles. Finally, thermal conductance, and thermal conductivity were considered, as microscopic observations of heat releases are related to mesoscopic hydrodynamic processes in the cell governed by those thermal transportation coefficients. These quantities have been widely measured for inorganic materials, while similar techniques are not immediately applicable to biological molecules with inhomogeneity in composition and shape such as proteins. Several simulation techniques have been proposed, although each technique can be improved for computational costs or a need for *a priori* knowledge of empirical parameters to run simulations.

As reviewed here, currently available simulation toolkits can be more extensively applied to realistic molecular systems, such as myosin and SERCA, which are both essential players in thermogenic processes in living organisms. Simulation studies on biochemical reactions and rearrangements of biopolymer complexes may provide microscopic insights into heat release and its uses. Those contributions will enrich the possible scenarios of functional expressions of intracellular processes around the heat spots, to be followed by experimental studies when methods are available. Conversely, simulation techniques for the estimation of heat transfer coefficients need to be improved. Clarifying the thermal properties of biopolymers has indispensable requirements, as these values are often inaccessible with conventional experimental techniques. Reliable estimations of heat transfer coefficients are a theoretical basis to address the trans-scale thermal signaling from MD to hydrodynamics in the cell.

We anticipate that simulation techniques will be more sophisticated to be comparable with experimental observations in near future. Likewise, experimental techniques should become accessible for information on biological molecules at microscopic and atomistic scales. Developments of both techniques in concert are expected to achieve the goal, and finally to lead us to a comprehensive understanding of the novel roles of cellular thermogenesis as the trans-scale thermal signaling although these expectations have not been realized so far.

## Acknowledgments

This work was supported by JSPS KAKENHI Grant Numbers 22H05053 (to IK and MS) and 22K19273 (to MS), by Yamada Science Foundation (to MS), by the Mitsubishi Foundation, and by the Human Frontier Science Program RGP0047/2018 (to MS).

## References

[1]     Lowell BB, Spiegelman BM (2000) Towards a molecular understanding of adaptive thermogenesis. Nature 404: 652-660. DOI: 10.1038/35007527

[2]     Tseng YH, Cypess AM, Kahn CR (2010) Cellular bioenergetics as a target for obesity therapy. Nat Rev Drug Discov 9: 465-482. DOI: 10.1038/nrd3138

[3]     Oyama K, Ishii S, Suzuki M (2022) Opto-thermal technologies for microscopic analysis of cellular temperature-sensing systems. Biophys Rev 14: 41-54. DOI: 10.1007/s12551-021-00854-1

[4]     Oyama K, Zeeb V, Kawamura Y, Arai T, Gotoh M, et al. (2015) Triggering of high-speed neurite outgrowth using an optical microheater. Sci Rep-Uk 5. DOI: 10.1038/srep16611

[5]     Oyama K, Mizuno A, Shintani SA, Itoh H, Serizawa T, et al. (2012) Microscopic heat pulses induce contraction of cardiomyocytes without calcium transients. Biochem Biophys Res Commun 417: 607-612. DOI: 10.1016/j.bbrc.2011.12.015

[6]	Marino A, Arai S, Hou YY, Degl'Innocenti A, Cappello V, et al. (2017) Gold nanoshell-mediated remote myotube activation. ACS Nano 11: 2494-2508. DOI: 10.1021/acsnano.6b08202

[7]	Oyama K, Zeeb V, Yamazawa T, Kurebayashi N, Kobirumaki-Shimozawa F, et al. (2022) Heat-hypersensitive mutants of ryanodine receptor type 1 revealed by microscopic heating. Proc Natl Acad Sci U S A 119: e2201286119. DOI: 10.1073/pnas.2201286119

[8]	Alon U (2007) Network motifs: Theory and experimental approaches. Nat Rev Genet 8: 450-461. DOI: 10.1038/nrg2102

[9]	Han JDJ (2008) Understanding biological functions through molecular networks. Cell Res 18: 224-237. DOI: 10.1038/cr.2008.16

[10]	Dyla M, Kjærgaard M, Poulsen H, Nissen P (2020) Structure and mechanism of P-type ATPase ion pumps. Annu Rev Biochem 89: 583-603. DOI: 10.1146/annurev-biochem-010611-112801

[11]	Pearl LH (2016) Review: The HSP90 molecular chaperone—An enigmatic ATPase. Biopolymers 105: 594-607. DOI: 10.1002/bip.22835

[12]	Geeves MA (2016) Review: The ATPase mechanism of myosin and actomyosin. Biopolymers 105: 483-491. DOI: 10.1002/bip.22853

[13]	Leyva JA, Bianchet MA, Amzel LM (2003) Understanding ATP synthesis: Structure and mechanism of the F1-ATPase (Review) (Review). Mol Membr Biol 20: 27-33. DOI: 10.1080/0968768031000066532

[14] Maracci C, Rodnina MV (2016) Review: Translational GTPases. Biopolymers 105: 463-475. DOI: 10.1002/bip.22832

[15] Lu SY, Jang H, Gu S, Zhang J, Nussinov R (2016) Drugging Ras GTPase: A comprehensive mechanistic and signaling structural view. Chem Soc Rev 45: 4929-4952. DOI: 10.1039/c5cs00911a

[16] Tulub AA, Stefanov VE (2001) Cisplatin stops tubulin assembly into microtubules. A new insight into the mechanism of antitumor activity of platinum complexes. Int J Biol Macromol 28: 191-198. DOI: 10.1016/S0141-8130(00)00159-8

[17] Ross EM, Wilkie TM (2000) GTPase-activating proteins for heterotrimeric G proteins: Regulators of G protein signaling (RGS) and RGS-like proteins. Annu Rev Biochem 69: 795-827. DOI: 10.1146/annurev.biochem.69.1.795

[18] Mardirossian N, Mcclain JD, Chan GKL (2018) Lowering of the complexity of quantum chemistry methods by choice of representation. J Chem Phys 148: 044106. DOI: 10.1063/1.5007779

[19] Warshel A, Levitt M (1976) Theoretical studies of enzymic reactions: Dielectric, electrostatic and steric stabilization of the carbonium ion in the reaction of lysozyme. J Mol Biol 103: 227-249. DOI: 10.1016/0022-2836(76)90311-9

[20] Petersen L, Ardèvol A, Rovira C, Reilly PJ (2009) Mechanism of cellulose hydrolysis by inverting GH8 endoglucanases: A QM/MM metadynamics study. J Phys Chem B 113: 7331-7339. DOI: 10.1021/jp811470d

[21] Van Der Kamp MW, Mulholland AJ (2013) Combined quantum mechanics/molecular mechanics (QM/MM) methods in computational Enzymology. Biochemistry 52: 2708-2728. DOI: 10.1021/bi400215w

[22] Khrenova MG, Grigorenko BL, Kolomeisky AB, Nemukhin AV (2015) Hydrolysis of guanosine triphosphate (GTP) by the Ras center dot GAP Protein Complex: Reaction Mechanism and Kinetic Scheme. J Phys Chem B 119: 12838-12845. DOI: 10.1021/acs.jpcb.5b07238

[23] Recabarren R, Osorio EH, Caballero J, Tuñón I, Alzate-Morales JH (2019) Mechanistic insights into the phosphoryl transfer reaction in cyclin-dependent kinase 2: A QM/MM study. PLOS ONE 14: e0215793. DOI: 10.1371/journal.pone.0215793

[24] Recabarren R, Zinovjev K, Tuñón I, Alzate-Morales J (2021) How a second $Mg^{2+}$ ion affects the phosphoryl-transfer mechanism in a protein kinase: A computational study. ACS Catal 11: 169-183. DOI: 10.1021/acscatal.0c03304

[25] Kanematsu Y, Narita A, Oda T, Koike R, Ota M, et al. (2022) Structures and mechanisms of actin ATP hydrolysis. Proc Natl Acad Sci U S A 119: e2122641119. DOI: 10.1073/pnas.2122641119

[26] Grigorenko BL, Rogov AV, Topol IA, Burt SK, Martinez HM, et al. (2007) Mechanism of the myosin catalyzed hydrolysis of ATP as rationalized by molecular modeling. Proc Natl Acad Sci U S A 104: 7057-7061. DOI: 10.1073/pnas.0701727104

[27] Grigorenko BL, Kaliman IA, Nemukhin AV (2011) Minimum energy reaction profiles for ATP hydrolysis in myosin. J Mol Graph Model 31: 1-4. DOI: 10.1016/j.jmgm.2011.07.005

[28]     Wickstrand C, Katona G, Nakane T, Nogly P, Standfuss J, et al. (2020) A tool for visualizing protein motions in time-resolved crystallography. Struct Dyn 7: 024701. DOI:10.1063/1.5126921

[29]     Tran V, Karsai A, Fong MC, Cai WL, Fraley JG, et al. (2020) Direct visualization of the binding of transforming growth factor beta 1 with cartilage oligomeric matrix protein via high-resolution atomic force microscopy. J Phys Chem B 124: 9497-9504. DOI: 10.1021/acs.jpcb.0c07286

[30]     Nakajima K, Yamazaki T, Kimura Y, So M, Goto Y, et al. (2020) Time-resolved observation of evolution of amyloid-beta oligomer with temporary salt crystals. J Phys Chem Lett 11: 6176-6184. DOI: 10.1021/acs.jpclett.0c01487

[31]     Mizutani Y, Mizuno M (2022) Time-resolved spectroscopic mapping of vibrational energy flow in proteins: Understanding thermal diffusion at the nanoscale. Journal of Chemical Physics 157, 240901. DOI: 10.1063/5.0116734

[32]     Kurisaki I, Tanaka S (2021) Elucidating microscopic events driven by GTP hydrolysis reaction in the Ras-GAP system with semi-reactive molecular dynamics simulations: The alternative role of a phosphate binding loop for mechanical energy storage. Phys Chem Chem Phys 23: 26151-26164. DOI: 10.1039/d1cp04061h

[33]     Takayanagi M, Nagaoka M (2011) Incipient structural and vibrational relaxation process of photolyzed carbonmonoxy myoglobin: Statistical analysis by perturbation ensemble molecular dynamics method. Theor Chem Acc 130: 1115-1129. DOI: 10.1007/s00214-011-0992-y

[34]     Takayanagi M, Okumura H, Nagaoka M (2007) Anisotropic structural relaxation and its correlation with the excess energy diffusion in the incipient process of photodissociated MbCO: High-

resolution analysis via ensemble perturbation method. J Phys Chem B 111: 864-869. DOI: 10.1021/jp066340+

[35]     Ross J (2006) Energy transfer from adenosine triphosphate. J Phys Chem B 110: 6987-6990. DOI: 10.1021/jp0556862

[36]     Ogata K, Shen JW, Sugawa S, Nakamura S (2012) MD Simulation Study of Ras/Raf Dissociation and the Resonating Structure of Deactivated Ras. B Chem Soc Jpn 85: 1318-1328. DOI: 10.1246/bcsj.20120065

[37]     Dignon GL, Best RB, Mittal J (2020) Biomolecular phase separation: From molecular driving forces to macroscopic properties. Annu Rev Phys Chem 71: 53-75. DOI: 10.1146/annurev-physchem-071819-113553

[38]     Gomes E, Shorter J (2019) The molecular language of membraneless organelles. J Biol Chem 294: 7115-7127. DOI: 10.1074/jbc.TM118.001192

[39]     Hofmann S, Kedersha N, Anderson P, Ivanov P (2021) Molecular mechanisms of stress granule assembly and disassembly. Biochim Biophys Acta Mol Cell Res 1868: 118876. DOI: 10.1016/j.bbamcr.2020.118876

[40]     Laplante M, Sabatini DM (2012) MTOR signaling in growth control and disease. Cell 149: 274-293. DOI: 10.1016/j.cell.2012.03.017

[41]     Saxton RA, Sabatini DM (2017) MTOR signaling in growth, metabolism, and disease. Cell 168: 960-976. DOI: 10.1016/j.cell.2017.02.004

[42]     Shen K, Rogala KB, Chou HT, Huang RK, Yu ZH, et al. (2019) Cryo-EM structure of the human FLCN-FNIP2-Rag-Ragulator complex. Cell 179: 1319-1329.e8+. DOI: 10.1016/j.cell.2019.10.036

[43]     Lans H, Hoeijmakers JHJ, Vermeulen W, Marteijn JA (2019) The DNA damage response to transcription stress. Nat Rev Mol Cell Biol 20: 766-784. DOI: 10.1038/s41580-019-0169-4

[44]     Sancar A, Lindsey-Boltz LA, Unsal-Kaçmaz K, Linn S (2004) Molecular mechanisms of mammalian DNA repair and the DNA damage checkpoints. Annu Rev Biochem 73: 39-85. DOI: 10.1146/annurev.biochem.73.011303.073723

[45]     Bohnsack KE, Bohnsack MT (2019) Uncovering the assembly pathway of human ribosomes and its emerging links to disease. EMBO J 38: e100278. DOI: 10.15252/embj.2018100278

[46]     Kaczanowska M, Rydén-Aulin M (2007) Ribosome biogenesis and the translation process in Escherichia coli. Microbiol Mol Biol Rev 71: 477-494. DOI: 10.1128/MMBR.00013-07

[47]     Borišek J, Casalino L, Saltalamacchia A, Mays SG, Malcovati L, et al. (2021) Atomic-level mechanism of Pre-mRNA splicing in health and disease. Acc Chem Res 54: 144-154. DOI: 10.1021/acs.accounts.0c00578

[48]     Fica SM, Nagai K (2017) Cryo-electron microscopy snapshots of the spliceosome: Structural insights into a dynamic ribonucleoprotein machine. Nat Struct Mol Biol 24: 791-799. DOI: 10.1038/nsmb.3463

[49]     Shi YG (2017) Mechanistic insights into precursor messenger RNA splicing by the spliceosome. Nat Rev Mol Cell Biol 18: 655-670. DOI: 10.1038/nrm.2017.86

[50]     Heck AJR (2008) Native mass spectrometry: A bridge between interactomics and structural biology. Nat Methods 5: 927-933. DOI: 10.1038/nmeth.1265

[51]     Levy ED, Erba EB, Robinson CV, Teichmann SA (2008) Assembly reflects evolution of protein complexes. Nature 453: 1262-1265-U1266. DOI: 10.1038/nature06942

[52]     Strunk BS, Karbstein K (2009) Powering through ribosome assembly. Rna 15: 2083-2104. DOI: 10.1261/rna.1792109

[53]     Hall Z, Politis A, Bush MF, Smith LJ, Robinson CV (2012) Charge-state dependent compaction and dissociation of protein complexes: Insights from ion mobility and Molecular Dynamics. J Am Chem Soc 134: 3429-3438. DOI: 10.1021/ja2096859

[54]     Hall Z, Hernández H, Marsh JA, Teichmann SA, Robinson CV (2013) The role of salt bridges, charge density, and subunit flexibility in determining disassembly routes of protein complexes. Structure 21: 1325-1337. DOI: 10.1016/j.str.2013.06.004

[55]     Marsh JA, Hernández H, Hall Z, Ahnert SE, Perica T, et al. (2013) Protein complexes are under evolutionary selection to assemble via ordered pathways. Cell 153: 461-470. DOI: 10.1016/j.cell.2013.02.044

[56]     Boeri Erba E, Petosa C (2015) The emerging role of native mass spectrometry in characterizing the structure and dynamics of macromolecular complexes. Protein Sci 24: 1176-1192. DOI: 10.1002/pro.2661

[57] Lössl P, Van De Waterbeemd M, Heck AJR (2016) The diverse and expanding role of mass spectrometry in structural and molecular biology. EMBO J 35: 2634-2657. DOI: 10.15252/embj.201694818

[58] Van De Waterbeemd M, Tamara S, Fort KL, Damoc E, Franc V, et al. (2018) Dissecting ribosomal particles throughout the kingdoms of life using advanced hybrid mass spectrometry methods. Nat Commun 9: 2493. DOI: 10.1038/s41467-018-04853-x

[59] Boeri Erba E, Signor L, Petosa C (2020) Exploring the structure and dynamics of macromolecular complexes by native mass spectrometry. J Proteomics 222: 103799. DOI: 10.1016/j.jprot.2020.103799

[60] Ahsan N, Rao RSP, Wilson RS, Punyamurtula U, Salvato F, et al. (2021) Mass spectrometry-based proteomic platforms for better understanding of SARS-CoV-2 induced pathogenesis and potential diagnostic approaches. Proteomics 21: e2000279. DOI: 10.1002/pmic.202000279

[61] Lensink MF, Brysbaert G, Mauri T, Nadzirin N, Velankar S, et al. (2021) Prediction of protein assemblies, the next frontier: The CASP14-Capri experiment. Proteins 89: 1800-1823. DOI: 10.1002/prot.26222

[62] Soni N, Madhusudhan MS (2017) Computational modeling of protein assemblies. Curr Opin Struct Biol 44: 179-189. DOI: 10.1016/j.sbi.2017.04.006

[63] Peterson LX, Togawa Y, Esquivel-Rodriguez J, Terashi G, Christoffer C, et al. (2018) Modeling the assembly order of multimeric heteroprotein complexes. PLOS Comput Biol 14:

e1005937. DOI: 10.1371/journal.pcbi.1005937

[64]    Aderinwale T, Christoffer C, Kihara D (2022) RL-MLZerD: Multimeric protein docking using reinforcement learning. Front Mol Biosci 9: 969394. DOI: 10.3389/fmolb.2022.969394

[65]    Patel A, Malinovska L, Saha S, Wang J, Alberti S, et al. (2017) ATP as a biological hydrotrope. Science 356: 753-756. DOI: 10.1126/science.aaf6846

[66]    Sumino A, Yamamoto D, Iwamoto M, Dewa T, Oiki S (2014) Gating-associated clustering-dispersion dynamics of the KcsA potassium channel in a lipid membrane. J Phys Chem Lett 5: 578-584. DOI: 10.1021/jz402491t

[67]    Kurisaki I, Tanaka S (2021) Reaction pathway sampling and free-energy analyses for multimeric protein complex disassembly by employing hybrid configuration bias Monte Carlo/Molecular Dynamics simulation. ACS Omega 6: 4749-4758. DOI: 10.1021/acsomega.0c05579

[68]    Aquilina JA, Robinson CV (2003) Investigating interactions of the pentraxins serum amyloid P component and C-reactive protein by mass spectrometry. Biochem J 375: 323-328. DOI: 10.1042/Bj20030541

[69]    Kurisaki I, Tanaka S (2022) Computational prediction of heteromeric protein complex disassembly order using hybrid Monte Carlo/molecular dynamics simulation. Phys Chem Chem Phys 24: 10575-10587. DOI: 10.1039/d2cp00267a

[70]    Zwier MC, Adelman JL, Kaus JW, Pratt AJ, Wong KF, et al. (2015) WESTPA: An interoperable, highly scalable software package for weighted ensemble simulation and analysis. J Chem Theory Comput 11: 800-809. DOI: 10.1021/ct5010615

[71]     Zwier MC, Pratt AJ, Adelman JL, Kaus JW, Zuckerman DM, et al. (2016) Efficient atomistic simulation of pathways and calculation of rate constants for a protein-peptide binding process: Application to the MDM2 protein and an intrinsically disordered p53 peptide. J Phys Chem Lett 7: 3440-3445. DOI: 10.1021/acs.jpclett.6b01502

[72]     Zuckerman DM, Chong LT (2017) Weighted ensemble simulation: Review of methodology, applications, and software. Annu Rev Biophys 46: 43-57. DOI: 10.1146/annurev-biophys-070816-033834

[73]     Harada R (2018) Simple, yet efficient conformational sampling methods for reproducing/predicting biologically rare events of proteins. B Chem Soc Jpn. 91: 1436-1450. DOI: 10.1246/bcsj.20180170

[74]     Moritsugu K, Ekimoto T, Ikeguchi M, Kidera A (2023) Binding and unbinding pathways of peptide substrates on the SARS-CoV-2 3CL protease. J Chem Inf Model 63: 240-250. DOI: 10.1021/acs.jcim.2c00946

[75]     Tran DP, Kitao A (2019) Dissociation process of a MDM2/p53 complex investigated by parallel cascade selection Molecular Dynamics and the markov state model. J Phys Chem B 123: 2469-2478. DOI: 10.1021/acs.jpcb.8b10309

[76]     Howard R, Scheiner A, Cunningham J, Gatenby R (2019) Cytoplasmic convection currents and intracellular temperature gradients. PLOS Comput Biol 15: e1007372. DOI: 10.1371/journal.pcbi.1007372

[77]     Ellis RJ (2001) Macromolecular crowding: An important but neglected aspect of the intracellular environment. Curr Opin Struct Biol 11: 114-119. DOI: 10.1016/S0959-440x(00)00172-X

[78]     Bastos ARN, Brites CDS, Rojas-Gutierrez PA, Ferreira RAS, Longo RL, et al. (2020) Thermal properties of lipid bilayers derived from the transient heating regime of upconverting nanoparticles. Nanoscale 12: 24169-24176. DOI: 10.1039/d0nr06989b

[79]     Song P, Gao H, Gao ZS, Liu JX, Zhang RP, et al. (2021) Heat transfer and thermoregulation within single cells revealed by transient plasmonic imaging. Chem 7: 1569-1587. DOI: 10.1016/j.chempr.2021.02.027

[80]     Sotoma S, Zhong CX, Kah JCY, Yamashita H, Plakhotnik T, et al. (2021) In situ measurements of intracellular thermal conductivity using heater-thermometer hybrid diamond nanosensors. Sci Adv 7. DOI: 10.1126/sciadv.abd7888

[81]     Nakano T, Ohara T, Kikugawa G (2008) Study on molecular thermal energy transfer in a lipid bilayer. J Therm Sci Tech-Jpn. 3: 421-429. DOI: 10.1299/jtst.3.421

[82]     Nakano T, Kikugawa G, Ohara T (2010) A molecular dynamics study on heat conduction characteristics in DPPC lipid bilayer. J Chem Phys 133: 154705. DOI: 10.1063/1.3481650

[83]     Youssefian S, Rahbar N, Lambert CR, Van Dessel S (2017) Variation of thermal conductivity of DPPC lipid bilayer membranes around the phase transition temperature. J R Soc Interface 14. DOI: 10.1098/rsif.2017.0127

[84]     Heimburg T (2019). Phase transitions in biological membranes. Thermodynamics and biophysics of biomedical nanosystems, springer nature Singapore pte Ltd. DOI 10.1007/978-981-13-0989-2_3

[85]     Youssefian S, Rahbar N, Van Dessel S (2018) Thermal conductivity and rectification in asymmetric archaeal lipid membranes. J Chem Phys 148: 174901. DOI: 10.1063/1.5018589

[86]     Drolle E, Negoda A, Hammond K, Pavlov E, Leonenko Z (2017) Changes in lipid membranes may trigger amyloid toxicity in Alzheimer's disease. PLOS ONE 12: e0182194. DOI: ARTN e018219410.1371/journal.pone.0182194

[87]     Harayama T, Riezman H (2018) Understanding the diversity of membrane lipid composition. Nat Rev Mol Cell Biol 19: 281-296. DOI: 10.1038/nrm.2017.138

[88]     Ingólfsson HI, Carpenter TS, Bhatia H, Bremer PT, Marrink SJ, et al. (2017) Computational lipidomics of the neuronal plasma membrane. Biophys J 113: 2271-2280. DOI: 10.1016/j.bpj.2017.10.017

[89]     Grassi S, Giussani P, Mauri L, Prioni S, Sonnino S, et al. (2020) Lipid rafts and neurodegeneration: Structural and functional roles in physiologic aging and neurodegenerative diseases. J Lipid Res 61: 636-654. DOI: 10.1194/jlr.TR119000427

[90]     Foley BM, Gorham CS, Duda JC, Cheaito R, Szwejkowski CJ, et al. (2014) Protein thermal conductivity measured in the solid state reveals anharmonic interactions of vibrations in a fractal structure. J Phys Chem Lett 5: 1077-1082. DOI: 10.1021/jz500174x

[91]     Leitner DM (2013) Thermal boundary conductance and thermal rectification in molecules. J Phys Chem B 117: 12820-12828. DOI: 10.1021/jp402012z

[92]     Pandey HD, Leitner DM (2017) Influence of thermalization on thermal conduction through molecular junctions: Computational study of PEG oligomers. J Chem Phys 147: 084701. DOI: 10.1063/1.4999411

[93]     Pandey HD, Leitner DM (2018) Small saccharides as a blanket around proteins: A computational study. J Phys Chem B 122: 7277-7285. DOI: 10.1021/acs.jpcb.8b04632

[94]     Yamato T, Wang TT, Sugiura W, Laprévote O, Katagiri T (2022) Computational Study on the Thermal Conductivity of a Protein. J Phys Chem B 126: 3029-3036. DOI: 10.1021/acs.jpcb.2c00958

[95]     Lervik A, Bresme F, Kjelstrup S, Bedeaux D, Miguel Rubi JM (2010) Heat transfer in protein-water interfaces. Phys Chem Chem Phys 12: 1610-1617. DOI: 10.1039/b918607g

[96]     Kurisaki I, Tanaka S, Mori I, Umegaki T, Mori Y, et al. (2023) Thermal conductivity and conductance of protein in aqueous solution: Effects of geometrical shape. J Comput Chem 44: 857-868. DOI: 10.1002/jcc.27048

[97]     Schuler LD, Daura X, Van Gunsteren WF (2001) An improved GROMOS96 force field for aliphatic hydrocarbons in the condensed phase. J Comput Chem 22: 1205-1218. DOI: 10.1002/jcc.1078